\documentclass[aps,pre,preprint,unsortedaddress,letterpaper]{revtex4-1}

\usepackage{graphicx}
\usepackage{color}

\begin{document}

\title{Surface adsorption of lattice HP proteins: Thermodynamics and structural transitions using Wang-Landau sampling}

\author{Ying Wai Li}
\email[]{ywli@physast.uga.edu}
\affiliation{Center for Simulational Physics, University of Georgia, Athens, Georgia 30602, U.S.A.}

\author{Thomas W\"ust}
\affiliation{Swiss Federal Research Institute WSL, Z\"urcherstrasse 111, CH-8903 Birmensdorf, Switzerland}

\author{David P. Landau}
\affiliation{Center for Simulational Physics, University of Georgia, Athens, Georgia 30602, U.S.A.}

\date{\today}


\begin{abstract}
Wang-Landau sampling has been applied to investigate the thermodynamics and structural properties 
of a lattice hydrophobic-polar heteropolymer (the HP protein model) interacting with an attractive 
substrate. For simplicity, we consider a short HP sequence consisting of only 36 monomers 
interacting with a substrate which attracts all monomers in the sequence. The conformational 
``phase transitions'' have been identified by a canonical analysis of the 
specific heat and suitable structural observables. Three major ``transitions'', namely, adsorption, 
hydrophobic core formation and ``flattening'' of adsorbed structures, are observed. Depending on 
the surface attractive strength relative to the intra-protein attraction among the H monomers, 
these processes take place in different sequences upon cooling.
\end{abstract}

\maketitle


\section{Introduction}

Protein folding and protein adsorption have long been subjects of 
intense research. The topics gained so much attention both because of
their numerous applications in nanotechnology, biomaterials, medical
and biological sciences (see e.g., \cite{Horbett1995,Sarikaya2003,hlady}
and references therein), but also because of the many interesting, yet 
challenging, basic scientific questions they pose. However, due to the 
complexity of the problems, our understanding of protein structure and 
folding is still incomplete and a ``general theory'' is lacking. The 
difficulties arise from the many interactions, of various strengths, 
among the building blocks constituting the protein organizing the 
molecules into primary, secondary or tertiary structures \cite{Branden}.

Additional protein-substrate interactions add a further level of
complexity to the problem. When brought in the vicinity of a
substrate, a protein rearranges its configuration to differ from
its native state. Generally, the nature of surface adsorption depends
on the properties of the protein, surface and interactions,
but many details remain unsolved \cite{hlady,Haynes1994,rabe}.

Simplified protein models that capture the essential features of real
proteins have thus been proposed in hope of unraveling the mysteries. 
Such coarse-grained models have the advantage of being readily 
accessible to computer simulation. With the emerging computer
power nowadays, numerical simulation of such protein models play
therefore an ever important role to the understanding of the problems. 
Nevertheless, they are still computationally intensive and unexpectedly 
difficult despite the simplicity of the models. In this work, we will 
investigate the hydrophobic-polar (HP) lattice protein model 
\cite{Dill1985} (one of the simplest such model), subjected to an attractive 
surface. Thereby, we attempt to understand the thermodynamic and 
\emph{generic} structural behavior of protein adsorption from a 
statistical mechanics point of view. Furthermore, we will study the 
influence of the strength of substrate attraction on the adsorption 
and folding processes of an HP protein.


\section{Model}
In an aqueous environment, a protein's hydrophobic amino acids tend to stay away from the solvent 
and form an interior core, while polar amino acids form an exterior shielding shell 
\cite{Kendrew1958,Kendrew1960}. This is known as the hydrophobic effect and is believed to be the 
driving force for the formation of tertiary structures \cite{Sturtevant1977,Baldwin1986,
Spolar1989}. The HP model \cite{Dill1985} is a prototypical, coarse-grained lattice protein model 
introduced to mimic this phenomenon. It classifies amino acids into two types of monomers 
according to their affinity to water: hydrophobic (H) and polar (P). Interactions are restricted 
to an attractive coupling, of strength $\varepsilon_{HH}$, between non-bonded hydrophobic monomers 
occupying nearest-neighbor sites. 

To simulate protein adsorption, an attractive substrate is placed at $z = 0$ on a 3-dimensional 
cubic lattice \cite{Bachmann2006}. For simplicity, in this work we consider a surface that 
attracts all monomers with a strength $\varepsilon_S$. The energy of the system is then given by:
\begin{equation}
  E = - \varepsilon_{HH} n_{HH} - \varepsilon_S n_S,
  \label{eq:hamiltonian}
\end{equation}

\begin{figure}[h!]
\includegraphics[width=14pc]{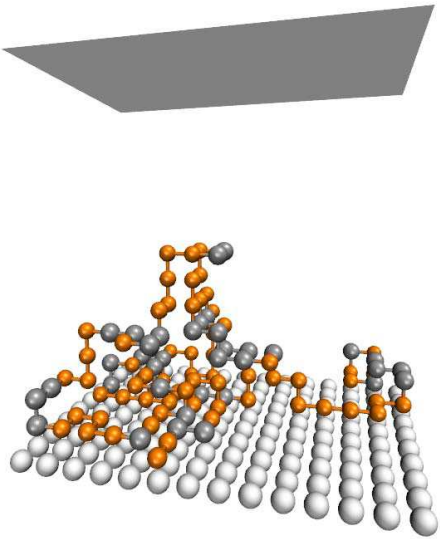}\hspace{2pc}%
\begin{minipage}[b]{21pc}\caption{\label{HPmodel}A schematic diagram showing the model used in this work. The gray spheres represent hydrophobic monomers, orange spheres represent polar monomers, faint 
spheres are the attractive molecules of the substrate and the solid top surface is non-attractive.}
\end{minipage}
\end{figure}

\noindent
$n_{HH}$ being the number of H-H interacting pairs and $n_S$ being the number of monomers 
adjacent to the bottom surface. In addition, a non-attractive wall is placed at $z = N + 1$ 
to confine the HP chain from above, where $N$ is the number of monomers in the sequence. The 
purpose of placing the steric upper wall is to limit the vertical translational degrees of 
freedom of the protein, so as to shorten the time spent on simulating desorbed conformations. 
Figure \ref{HPmodel} shows a schematic diagram of the model. Here, we have used a 36mer 
(P$_3$H$_2$P$_2$H$_2$P$_5$H$_7$P$_2$H$_2$P$_4$H$_2$P$_2$HP$_2$) \cite{{Unger1993}} to illustrate 
the effect of $\varepsilon_S$ on the structural ``phase transitions'' associated with protein 
adsorption.


\section{Methods}

\subsection{Wang-Landau sampling and production run}

In order to investigate such conformational transitions, it is necessary to have a mean of 
sampling a protein's entire conformational/energy space efficiently. Here, in the first stage 
of the simulation, Wang-Landau (WL) sampling \cite{WangLandau2001PRL,WangLandau2001PRE,
WangLandau2002,Landau2004} has been employed to estimate the energy density of states, $g(E)$, 
which then gives access to thermodynamic quantities at any temperature. In this iterative procedure, a trial configuration of energy $E_{trial}$ is generated with 
an acceptance probability inversely proportional to $g(E_{trial})$. The old configuration 
is retained if the trial one is rejected. A multiplicative modification factor $f$ (with an 
initial value $f_{init} = e^1$ at the beginning of the simulation) is used to modify $g(E)$, 
i.e., $g(E) \rightarrow g(E) \times f$. A histogram in energy is also accumulated: $H(E) 
\rightarrow H(E) + 1$. When a ``flat'' histogram is attained, the simulation is brought to 
the next iteration: $H(E)$ is reset and $f$ is reduced, $f \rightarrow \sqrt{f}$. A ``flat'' 
histogram is defined as when all entries in $H(E)$ are greater than $p \times H_{ave}$, 
where $p$ is the flatness criterion and $H_{ave}$ is the average of all entries in $H(E)$. 
All results presented in this work are obtained by using a flatness criterion $p = 0.8$ for 
reliable estimates of $g(E)$. The simulation is terminated when $\ln(f_{final})$ reaches a 
preset minimum value of $10^{-8}$.

The partition function, $Z(T)$ and subsequent thermodynamics, e.g. average energy 
$\left\langle E \right\rangle_T$ and specific heat $C_V(T)$, then follow:
\begin{equation}
Z(T) = \sum_{i}{g(E_i)e^{-E_i/kT}},
\label{eq:partitionFunction}
\end{equation}

\begin{equation}
\langle E \rangle_T = \frac{1}{Z(T)}\sum_{i}{E_ig(E_i)e^{-E_i/kT}},
\label{eq:energy}
\end{equation}

\begin{equation}
C_V(T) = \frac{1}{kT^2}\left(\langle E^2 \rangle - \left\langle E \right\rangle^2\right),
\label{eq:specificHeat}
\end{equation}

\noindent
where $k$ is the Boltzmann constant, $T$ is the temperature and the sum runs over all 
possible energies.

The second stage of the simulation consists in a production run making use of multicanonical sampling 
\cite{Berg1991,Berg1992} to generate a number of two-dimensional densities of states, $g(E,Q)$, where $Q$
is any structural quantity of interest, such as e.g., the radius of gyration $R_g = \sqrt{\frac{1}{N} 
\sum_i^N{\left(\vec{r}_i - \vec{r}_{cm}\right)^2}}$ ($\vec{r}_i$ is the position of the $i^{th}$ 
monomer and $\vec{r}_{cm}$ is the position of the center of mass), the number of surface contacts 
$n_S$, the number of surface contacts of hydrophobic monomer $n_{SH}$, the number of surface contacts of 
polar monomer $n_{SP}$, or the number of hydrophobic interaction pairs $n_{HH}$. During the production 
run, the inverse of $g(E)$, obtained from Wang-Landau sampling, is used as the weight of the acceptance rate, while structural quantities are calculated and two-dimensional histograms, $H(E,Q)$, are accumulated. 
At the end of the simulation, the $H(E,Q)$'s are reweighted by $g(E)$ in order to yield the two-dimensional densities of states, $g(E,Q)$'s. The partition function of an observable $Q$, $Z_Q(T)$, and its expectation value can then be calculated as
\begin{equation}
Z_Q(T) = \sum_{E,Q}{g(E,Q)e^{-E/kT}},
\label{ObservablePartitionFunction}
\end{equation}

\begin{equation}
\left\langle Q \right\rangle_T = \frac{1}{Z_Q(T)}\sum_{E,Q}{Q g(E,Q) e^{-E/kT}}.
\label{observable}
\end{equation}

Thermodynamics of the structural quantities in addition to the specific heat, $C_V$, are essential in identifying 
``transitions'' between different structural ``phases''. In cases where the specific heat 
shows ambiguous signals, structural quantities help clarifying the types of transition taking 
place at different temperatures. In some cases distinct signals might be missing in the specific heat,
whereas structural quantities are more reliable to identify structural transitions.

\subsection{Methodological pitfalls}

Traditional Monte Carlo trial moves for lattice polymers either change a conformation locally (e.g. 
kink flip and crankshaft) or non-locally (e.g. pivot moves). Local moves generate new configurations 
fairly similar to the old ones as most parts of the polymer remains unchanged, inducing long 
correlation times in the simulation. Non-local moves do not share the same problem, 
but they are ineffective for dense conformations. Two inventive Monte Carlo trial moves, namely, 
pull moves \cite{PullMoves} and bond-rebridging moves \cite{BRMoves}, have thus been implemented in our 
simulations to confront these problems. When combined with Wang-Landau sampling, 
they have proven to be particularly efficient in exploring conformational space \cite{Thomas2008,
WLSampling}. 

\begin{figure}[ht]
\includegraphics[width=18.8pc]{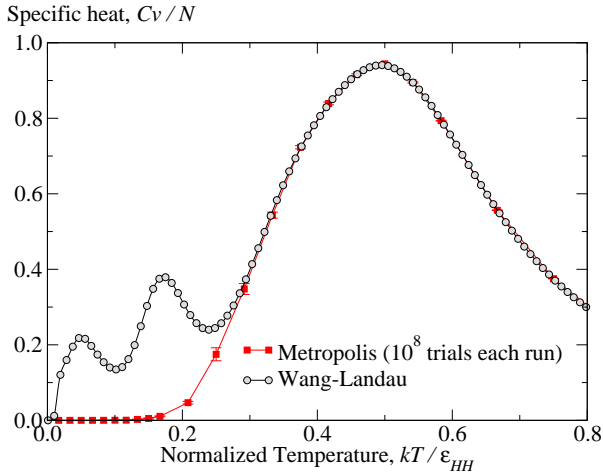}\hspace{2pc}%
\begin{minipage}[b]{17pc}\caption{\label{WLvsMetro} Comparison between Wang-Landau and Metropolis 
                                  sampling in obtaining the specific heat of a 36mer interacting 
                                  with a very weak attractive surface, in which $\varepsilon_{HH} = 
                                  12 \varepsilon_{S}$. Error bars smaller than the data points are 
                                  not shown.}
\end{minipage}
\end{figure}

However, these non-traditional trial moves alone are not able to give correct low temperature 
thermodynamics if they are used with Metropolis sampling, which is easily trapped in metastable 
states. Figure \ref{WLvsMetro} compares the two sampling methods, where the two 
transition peaks at low temperature in the specific heat are clearly missing in the Metropolis 
case. We thus stress that an appropriate combination of the sampling method and trial updates is 
crucial in obtaining correct results from a Monte Carlo simulation.
\vspace{3mm}


\section{Results}

\subsection{Identifying structural ``phase transitions''}

\begin{figure}[ht]
\centering
\includegraphics[width=0.45\columnwidth]{HP01HH12CvRg.eps}
\hspace{6mm}
\includegraphics[width=0.45\columnwidth]{HP01HH12SQ.eps} \\
\includegraphics[width=0.9\columnwidth]{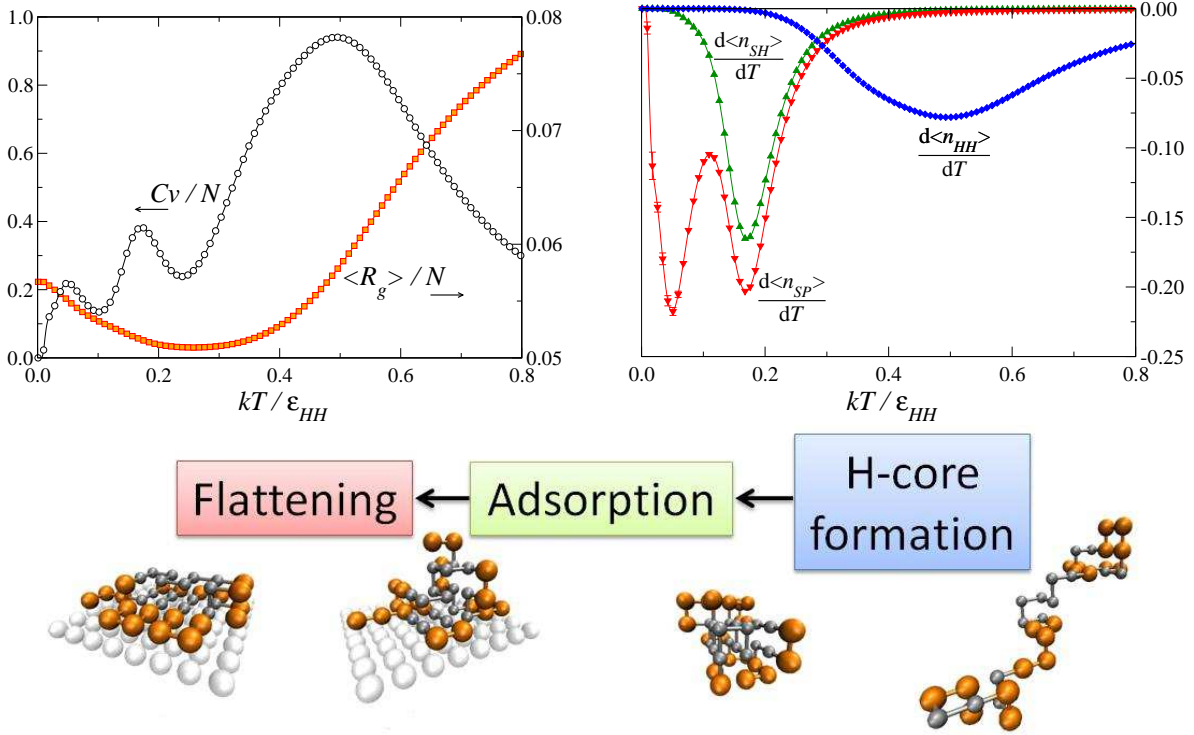}
\caption{\label{HP01HH12}Specific heat and structural quantities of the 36mer interacting with a very 
                         weak attractive surface ($\varepsilon_{HH} = 12 \varepsilon_{S}$). Error  
												 bars smaller than the data points are not shown.}
\end{figure}

The different structural ``phase transitions'' are best illustrated by considering the 36mer 
interacting with a very weak attractive surface ($\varepsilon_{HH} = 12\varepsilon_{S}$). Detailed 
studies of this system can be found in Refs.\cite{Li2011,Thomas2011}. As shown in Figure \ref{HP01HH12}, its specific 
heat has three distinct peaks, which represent three basic phase transitions, respectively (from high 
to low temperature): (i) hydrophobic core (H-core) formation, (ii) adsorption, and (iii) 
``flattening'' of the adsorbed structure.

These three transitions can be identified by comparing the 
peak positions of the specific heat and those of the structural quantities. The thermal derivative 
of $\left\langle n_{HH} \right\rangle$ peaks for H-core formation, while the thermal derivatives of 
$\left\langle n_{SH} \right\rangle$ and $\left\langle n_{SP} \right\rangle$ peak for adsorption at a 
higher temperature, and flattening at a lower temperature due to the fact that the flattening 
process has to take place after the protein is adsorbed on cooling. In some cases, it is also 
possible that flattening is signaled by a shoulder, or only a peak in either $\frac{d\left\langle n_{SH} \right\rangle}{dT}$ or $\frac{d\left\langle n_{SP} \right\rangle}{dT}$ but not both. 
\vspace{3mm}

\subsection{Effect of surface attraction on the sequence of transitions}

The three basic transitions occur at different temperatures when the surface attractive strength 
varies, giving rise to a \emph{different order} in structural changes. As a consequence, an extended, 
desorbed protein goes through a different path in conformational space towards the acquisition of 
compact, adsorbed ground states. Structures of the intermediate and ground states thus vary from case 
to case and are completely dependent on this sequence (or order) of transitions. 

\begin{figure}[ht]
\centering
\includegraphics[width=0.45\columnwidth]{HP01HH02CvRg.eps}
\hspace{6mm}
\includegraphics[width=0.45\columnwidth]{HP01HH02SQ.eps} \\
\includegraphics[width=0.9\columnwidth]{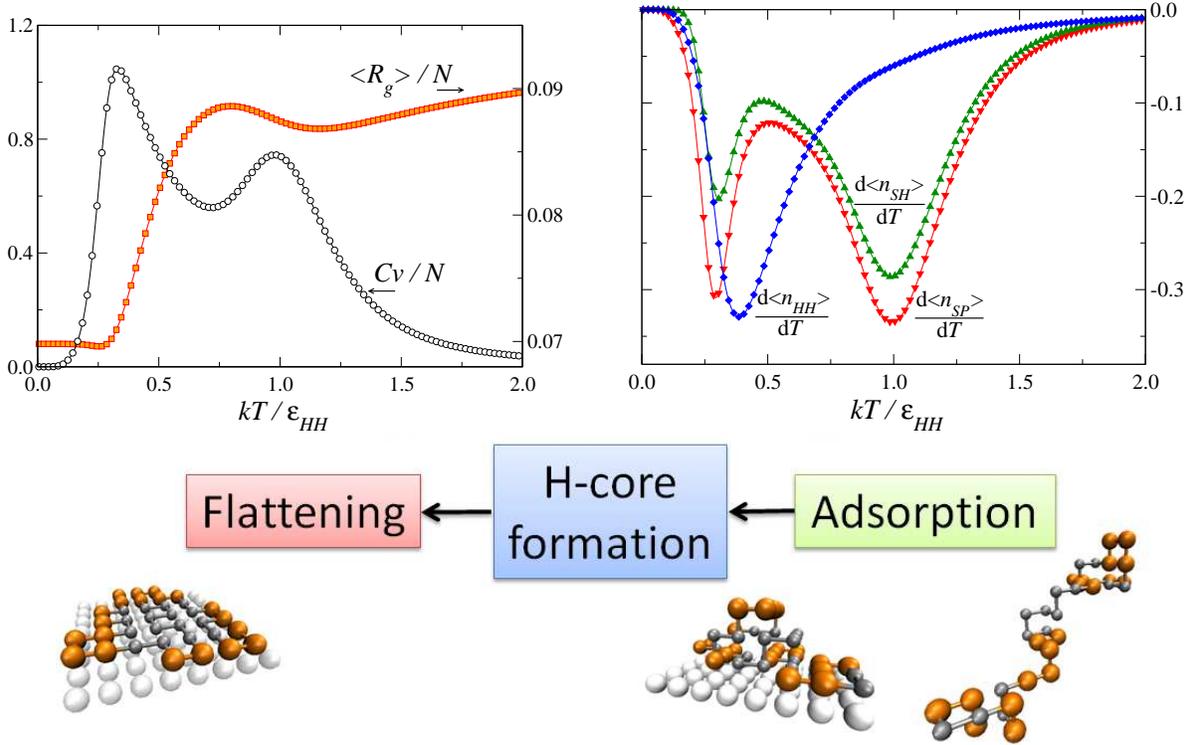}
\caption{\label{HP01HH02}Specific heat and structural quantities of the 36mer interacting with a 
                         moderately attractive surface ($\varepsilon_{HH} = 2 \varepsilon_{S}$).
												 Error bars are smaller than the data points.}
\end{figure}

When the surface attraction becomes stronger, it first affects the transitions at higher temperatures. 
Figure \ref{HP01HH02} shows the thermodynamics for the 36mer interacting with a surface of moderate 
attractive strength ($\varepsilon_{HH} = 2 \varepsilon_{S}$). Despite the fact that only two peaks 
are present in the specific heat, the three basic transitions and their order of occurrence are 
revealed by the analysis of the structural parameters. Adsorption takes place at $kT / 
\varepsilon_{HH} \approx 1.0$ corresponding to the $C_V$ peak at the higher temperature; a 
hydrophobic core forms at a slightly higher temperature ($kT / \varepsilon_{HH} \approx 0.35$) than that of flattening ($kT / \varepsilon_{HH} \approx 0.27$), and both processes are responsible for the $C_V$ peak at the 
lower temperature.


\vspace{3mm}

In terms of the transition sequence, H-core formation and adsorption have swapped places compared 
to the former case. On cooling, a three-dimensional, adsorbed but extended structure is first formed 
after adsorption which takes place at the highest temperature. The lowest energy state with a 
two-dimensional hydrophobic core is achieved after the combined action of H-core formation and 
flattening. As these two processes closely overlap, no intermediate states could be singled out between 
them.

\begin{figure}[ht]
\centering
\includegraphics[width=0.45\columnwidth]{HP02HH01CvRg.eps}
\hspace{6mm}
\includegraphics[width=0.45\columnwidth]{HP02HH01SQ.eps} \\
\includegraphics[width=0.9\columnwidth]{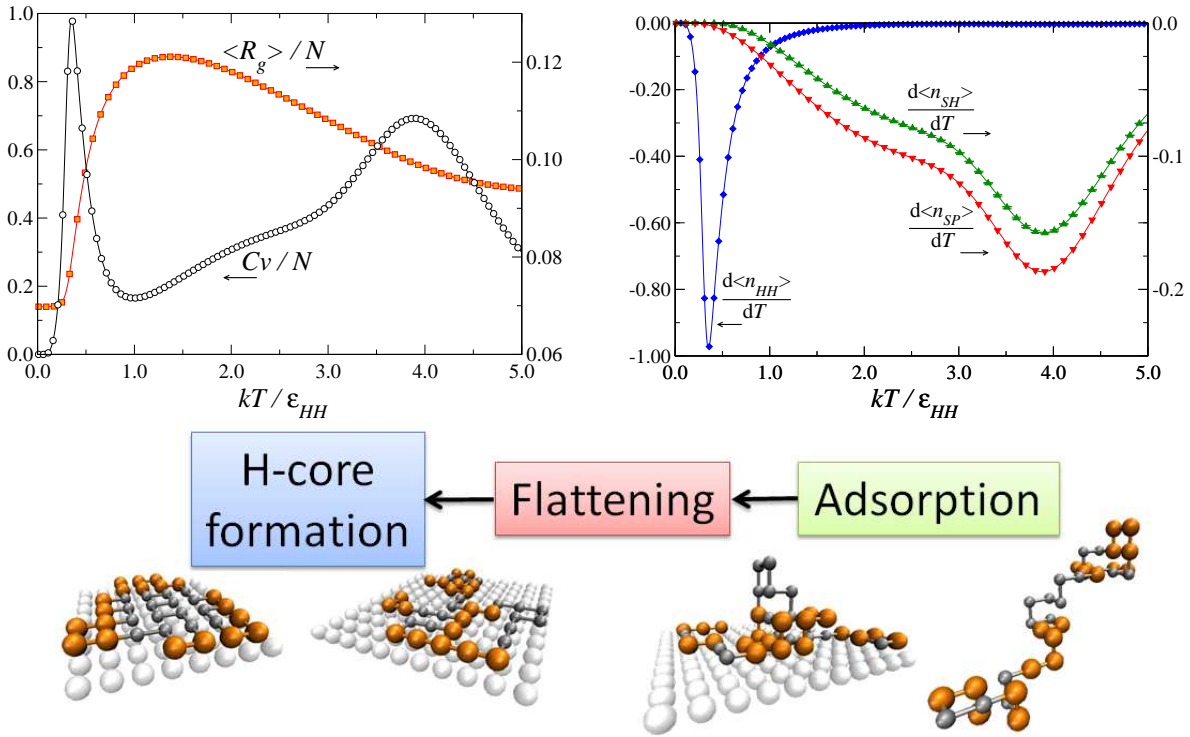}
\caption{\label{HP02HH01}Specific heat and structural quantities of the 36mer interacting with a 
                         strong attractive surface, in which $\varepsilon_{HH} = \frac{1}{2} 
												 \varepsilon_{S}$. Error bars are smaller than the data points.}
\end{figure}

Further increase in surface attractive strength shifts the H-core formation to an even lower temperature 
as shown in Figure \ref{HP02HH01}, where a strong attractive surface is used ($\varepsilon_{HH} = 
\frac{1}{2}\varepsilon_{S}$). In this case, there are also two peaks in the specific heat with a 
weak bump in between. A comparison with the structural properties clearly distinguishes the three 
basic transitions, which now occur at well separated temperatures. Adsorption 
takes place at $kT / \varepsilon_{HH} \approx 4.0$; H-core formation at $kT / \varepsilon_{HH} \approx 
0.4$; the bump occurs between $kT / \varepsilon_{HH} \approx 1.0$ and $kT / \varepsilon_{HH} \approx 
3.0$ is a signal of flattening.

With this transition ordering, the desorbed, extended protein first adsorbs on the surface to form a 
three-dimensional, adsorbed yet extended structure. After the flattening process, most of the monomers 
contact with the surface but the chain is still not compact. The H-core formation finally takes place 
on the surface, forming a two-dimensional ground state with a hydrophobic core.
\vspace{3mm}


\section{Conclusion}

Protein adsorption has been studied using the HP lattice model and Wang-Landau sampling with efficient Monte 
Carlo trial updates. The combined thermodynamic signals of the specific heat and some suitable structural quantities allowed us to identify conformational ``phase transitions''. 
Three basic transitions, namely, hydrophobic core formation, adsorption and ``flattening'' of adsorbed 
structures, have been found to occur in a different sequence with varying surface attraction strength, 
$\varepsilon_S$, upon cooling. The acquisition of the adsorbed, compact, lowest energy states from a 
desorbed, extended coil thus goes through different paths in conformational space with different 
intermediates. The structures of ground state configurations may also differ from each other for
different surface attractions.


\begin{acknowledgments}
This project is supported by the National Science Foundation (NSF) under grant no. DMR-0810223.
\end{acknowledgments}


\end{document}